\documentclass[12pt,preprint]{aastex}

\begin{document}


\title{Near Ultraviolet sources in the Great Observatories Origins Deep Survey Fields \\
    }


\author{D.F. de Mello\altaffilmark{1,2,3},  
T. Dahlen\altaffilmark{4}, Jonathan P. Gardner\altaffilmark{1}, N.A. Grogin\altaffilmark{3}}


\altaffiltext{1}{Laboratory for Observational Cosmology, Code 665, Goddard Space Flight Center, Greenbelt, MD
20771. duilia@ipanema.gsfc.nasa.gov}
\altaffiltext{2}{Catholic University of America Washington, DC 20064}

\altaffiltext{3}{Johns Hopkins University, Baltimore, MD 21218}

\altaffiltext{4}{Department of Physics, Stockholm University, SE-106 91 Stockholm, Sweden }


\begin{abstract}

We present an Ultraviolet (UV) selected sample of 268 objects in the two fields of the Great Observatories Origins 
Deep Survey (GOODS). We used the parallel observations 
taken with WFPC2 in the U--band (F300W) which covered 88\% of the GOODS fields to 
identify sources and selected only objects with GOODS/ACS counterparts. 
Spectroscopic redshifts for 95 of these sources are available and
we have used the multiwavelength GOODS data to estimate photometric redshifts
for the others. Most of the objects are between $0.2<z<0.8$. 
We used the spectral types obtained by the photometric redshift fitting 
to identify the starburst galaxies. We have also visually checked all 
objects and looked for tidal effects and nearby companions. We find that (i) 45\% of the UV-selected 
galaxies are starbursts, (ii) nearly 75\% of the starbursts have tidal tails or show some peculiarity 
typical of interactions or mergers, (iii) $\sim$50\% have companions 
within an area of 5$\times$5 arcsec. The UV-selected sample has an average 
rest-frame M$_{\rm B}$=--19.9 $\pm$ 0.1. The bluest objects in the sample ($U-B  < 0.2$ and $B-V < 0.1$) are at $1.1<z<1.9$ and have peculiar 
morphologies that resembles either tadpoles, chains, or double-clump galaxies.  Starbursts with tadpole or clump morphology at $z=0.8-1.3$ 
have sizes comparable to LBGs and compact Ultraviolet-luminous galaxies (UVLGs).

\end{abstract}


\section{Introduction}

One of the open questions in modern astronomy is to understand when 
galaxies acquired their morphology and how star formation and morphology
correlates. Are star-forming galaxies at higher$-z$, such as Lyman
Break galaxies (LBGs), related to local starburst galaxies or are there
evolutionary effects that prevent correlating the two populations? 
Heckman et al. (2005) attempted to identify the
local equivalents of LBGs using images from the
UV-satellite GALEX and spectroscopy from the SDSS. They found 
a class of nearby ($z<0.3$) UV luminous compact starburst galaxies (Compact UVLGs)  which resembles 
the LBGs at $z\sim 2-3$. No local galaxy population meets the UV luminous 
criteria which makes these objects a class of scaled-up unobscured starbursts just as
LBGs. If these objects are common out to $z<1-2$ they could contribute significantly 
to the rise of the star-formation rate density in the universe. 
Recently, Burgarella et al. (2006) using GALEX, Spitzer, HST and ground-based telescopes, selected a 
sample of $\sim$300 far-UV dropouts in the Chandra Deep Field South (CDF-S) which 
resembles LBGs at $z\sim 1$. Their sample contains not only UVLGs but also lower luminosity LBGs. 
However, they were able to obtain morphology for only 36 LBGs since only 1/4 of their GALEX field was 
observed with HST. In this article, we provide a sample of 268 near-UV sources detected in the two 
Great Observatories Origins Deep Survey (GOODS) fields for which there is extensive multiwavelength 
coverage. Our goal is to select enough star-forming galaxies at intermediate$-z$ 
in order to characterize their role in galaxy evolution. 

\section{The Data}

During the GOODS campaign with HST we implemented a WFPC2 pure parallel 
program aimed at covering the lack of U--band observations with ACS and
maximizing the synergy between the parallels and the prime instrument.
We took near-UV images with the F300W filter (U band with $\lambda_{\rm max}$ = 2920 \AA) 
in parallel of both fields from December 2002 until March 2003. GOODS/HST images, 
aiming also at searching for high$-z$ supernovae (Strolger et al. 2004), were taken 
in five repeat visits separated by 
approximately 45 days. The two GOODS fields were observed with the F435W (B), F606W (V), F775W (i), 
and F850LP (z) filters. The exposure times were 3, 2.5, 2.5 and 5 orbits per filter, respectively. 
The observations included the Hubble Deep Field North (HDF-N) and the CDF-S and each field had 15 ACS fields of view which makes GOODS 320
arcmin$^{2}$. 

\begin{figure}
\plotone{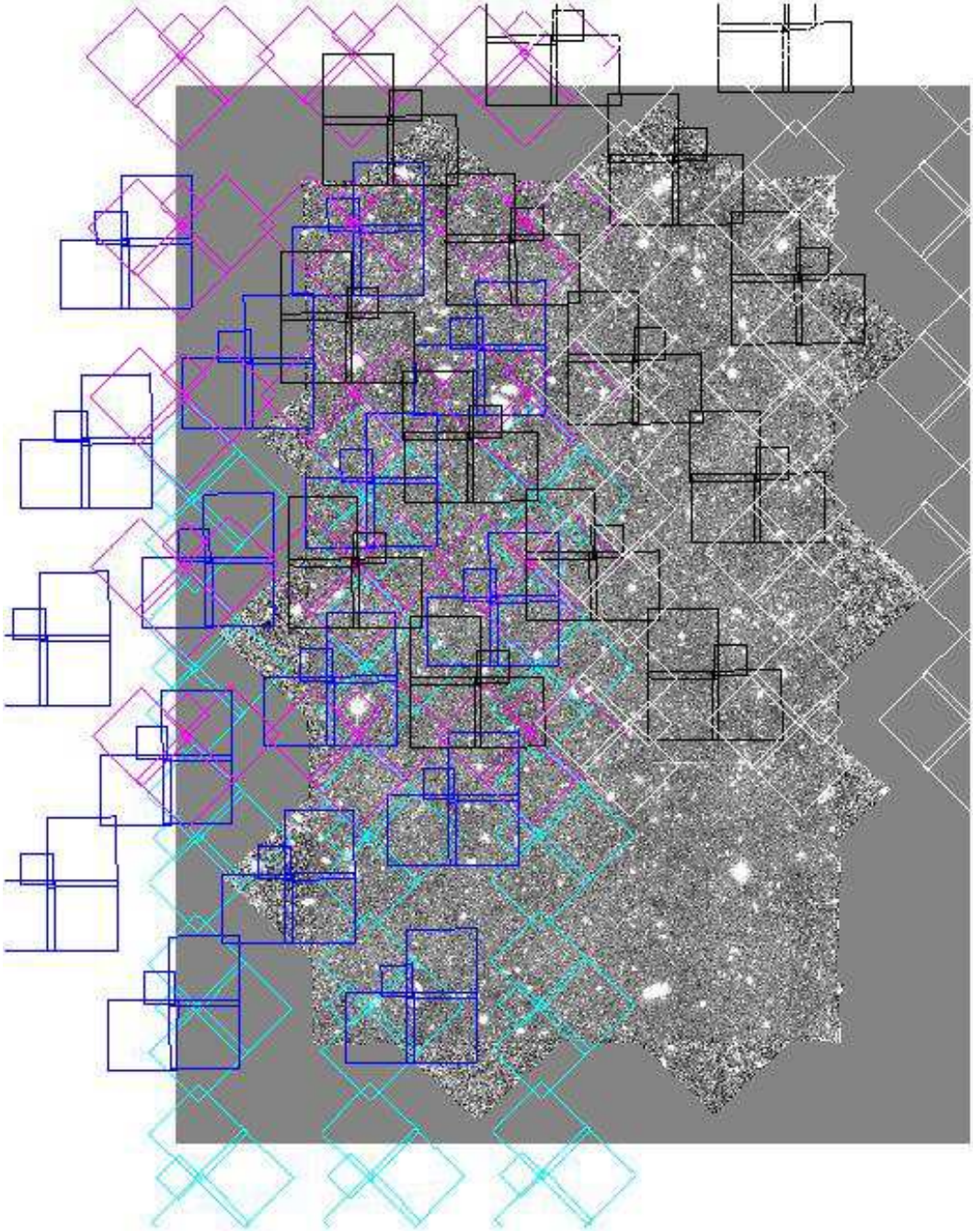}
\caption{WFPC2(F300W) coverage of the GOODS-S field. 
Each color corresponds to a different epoch. The ACS image is shown in the background. \label{allwfpc2frames}}
\end{figure}

\begin{figure}
\plotone{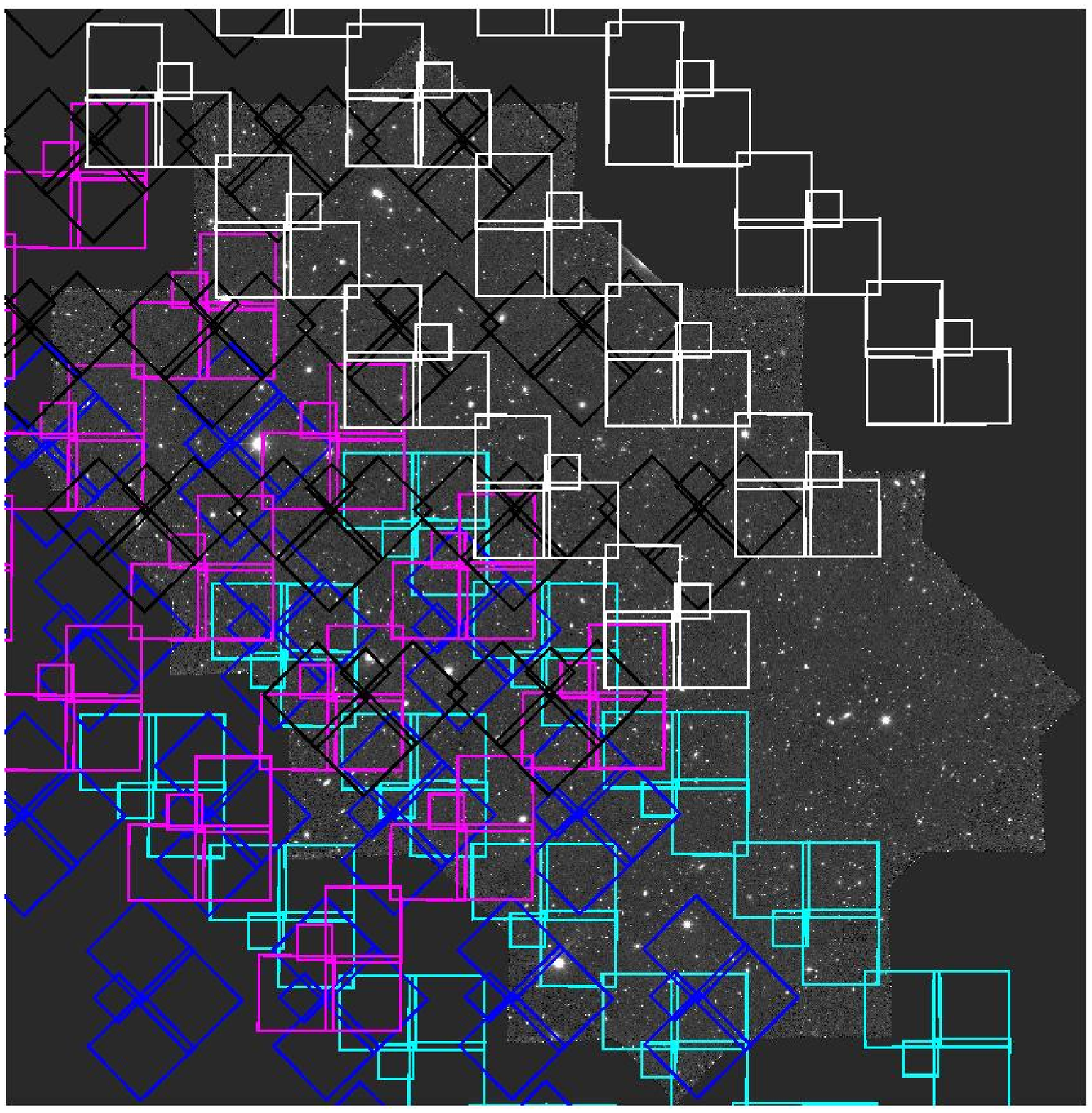}
\caption{WFPC2(F300W) coverage of the GOODS-N field. Each color corresponds to a different epoch. The ACS image 
is shown in the background.\label{hdfn_wfpc2_epall}}
\end{figure}

Due to the multi-epoch nature of the GOODS observations the 
WFPC2 coverage was not uniform and covered $\sim$ 88\% of the two fields. Fig.~\ref{allwfpc2frames}
and Fig.~\ref{hdfn_wfpc2_epall} show the overlap between the WFPC2 
observations and the ACS images.  

We retrieved a total of 741 WFPC2 images from the HST archive which were  
reduced using the HST pipeline.
We separated all images that were centered within $<$~0.1$^{'}$ from each 
other and dithered them using 
the package PyDrizzle which provides an automated method for
dither-combining and distortion-correcting images. The quality of
the drizzled image was checked by analyzing the PSF of a few stars
present in some of the images. 

A total of 30 WFPC2 drizzled images of the GOODS-South field and 25 
of the GOODS-North field were produced. The exposure times of each drizzled 
image varied from 800s to 11,100s with a typical exposure time being $\sim$2,000s. 
The deep mosaic with 11,100s falls mostly outside the GOODS-N field.

Throughout this paper, we use a cosmology
with $\Omega_{\rm M}=0.3$, $\Omega_{\Lambda}=0.7$~and $h=0.7$.
Magnitudes are given in the AB-system.

\section{Sample}

We detected sources on the drizzled images using SExtractor v2.2.2
(Bertin \& Arnouts 1996, hereafter SE). Our detection criterion was that a
source must exceed a $1.5\sigma$ sky threshold in 5 contiguous
pixels and we also used a detection filter with a Gaussian FWHM of
4 pixels. We used SE's MAG$_{-}$AUTO, which is calculated using a flexible elliptical
aperture around every detected object, and obtained the magnitude error, $\Delta$m, using SE's 
RMS calculated with BACKGROUND$_{-}$RMS.

\begin{figure}
\plotone{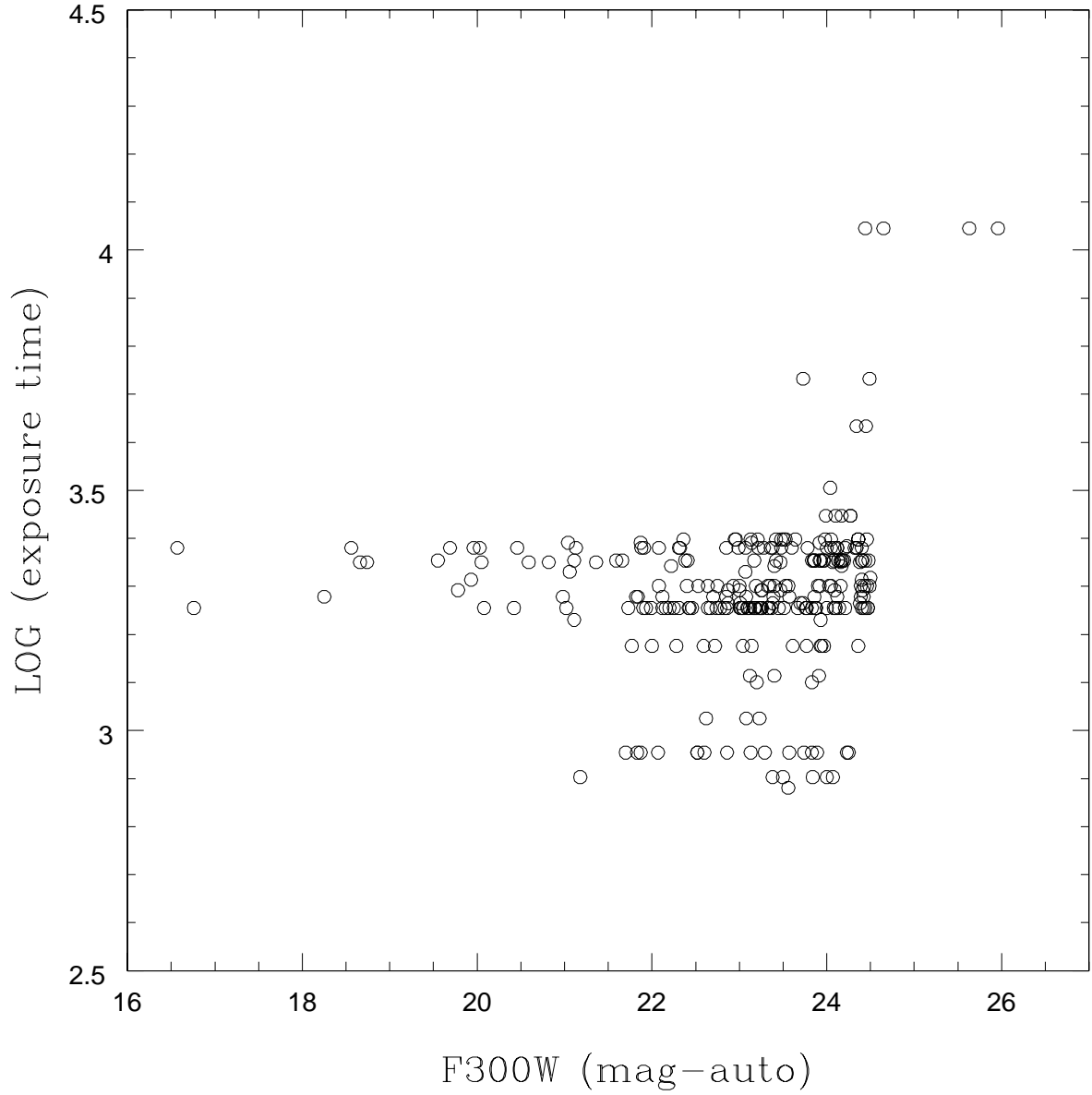}
\caption{The distribution of F300W magnitudes (mag$_{-}$auto(AB)) as a function of exposure time 
(seconds). The imposed magnitude limit of the sample is m=24.5. Two exceptions were made to the magnitude 
limit in order to include the two objects which are in the border of the deepest F300W mosaic 
(mag$_{-}$auto=25.63 and 25.96). \label{times}}
\end{figure}

The next step was to match the WFPC2 U--band coordinates with the GOODS ACS detections. The GOODS
catalog is z--band based with matched aperture photometry in the ACS B, V, i bands. We adopted a maximum
offset radius of 1.5 arcsec between the WFPC2 and ACS coordinates. In this paper we analyze only
the U--band selected objects which have ACS counterparts. Due to the shallowness and the 
heterogeneous nature of the parallel data we decided to focus on the bright-end 
by imposing a conservative magnitude limit to the sample. The magnitude limit 
mag$_{-}$auto~$<$~24.5 and error $\Delta$m$_{\rm AB}$(F300W)~$<$~0.10 were set to 
guarantee a clear identification even in the shallowest F300W images. The final
catalog has 130 objects in the GOODS-S and 138 in the GOODS-N. We have excluded duplicated 
objects that were in the borders of the drizzled images and were present in more than one field by choosing
the one with either higher exposure time or lower magnitude error. 
Fig.\ref{times} shows the distribution of magnitudes as a function of exposure times. Despite of the 
imposed magnitude limit at 24.5 we included two faint 
objects with magnitudes lower than the limit (mag$_{-}$auto=25.63 and 25.96) since they are in
the deepest F300W image and the $\Delta$m$_{\rm AB}$(F300W) are $<$~0.10.

We have identified 42 objects as stars based on their surface brightness versus magnitude relation. A total
of 32 are in the southern field and 10 in the northern field. Some of these objects 
are Active Galactic Nuclei (AGN) but are not further analyzed in this paper.

For comparison, the deepest near-UV image, obtained with 323.1 ks of 
HST time as part of the parallel observations of the Hubble Ultra
Deep Field campaign (de Mello et al. 2006), has 415 objects down to m$_{\rm AB}$(F300W)=27.5, 
which is 0.5 magnitudes deeper than the 45 HST/WFPC2(F300W) orbits dedicated to the
original HDF-N.

We have searched for differences in the magnitude distributions of the northern and southern fields 
that could be present due to cosmic variance (Somerville et al.
2004) and found no significant difference between the two fields. 
The Kolmogorov-Smirnov (KS) statistic gives less than 10\% confidence that these two fields
are drawn from different magnitude distributions, if we discard the stars on both samples. 
The fact that the southern field contains $\approx$ three times more stars than the northern field, 
shifts the average values of the southern sample towards brighter magnitudes and lowers the KS confidence 
level. 

\section{Photometric Redshifts, Spectral Types}

Photometric redshifts were calculated using the template fitting 
method described in detail by Dahlen et al. (2005).  The template SEDs 
used cover spectral types E, Sbc, Scd and Im (Coleman et al.  1980, 
with extension into UV and NIR-bands
by Bolzonella et al. 2000), and starburst templates number 2 and 3 from Kinney et al. (1996). 
We use multi-band photometry
for the GOODS fields, from $U$~to $K_s$~bands, obtained with both
$HST$~and ground-based facilities (Giavalisco et al. 2004). 
The WFPC2 data were not included in the photometric
redshift estimation. 
The redshift accuracy (Dahlen et al. 2005) of the GOODS method is 
$\Delta_z\equiv\langle|z_{\rm phot}-z_{\rm spec}|/(1+z_{\rm spec})\rangle\lesssim 0.1$. 

\begin{figure}
\plotone{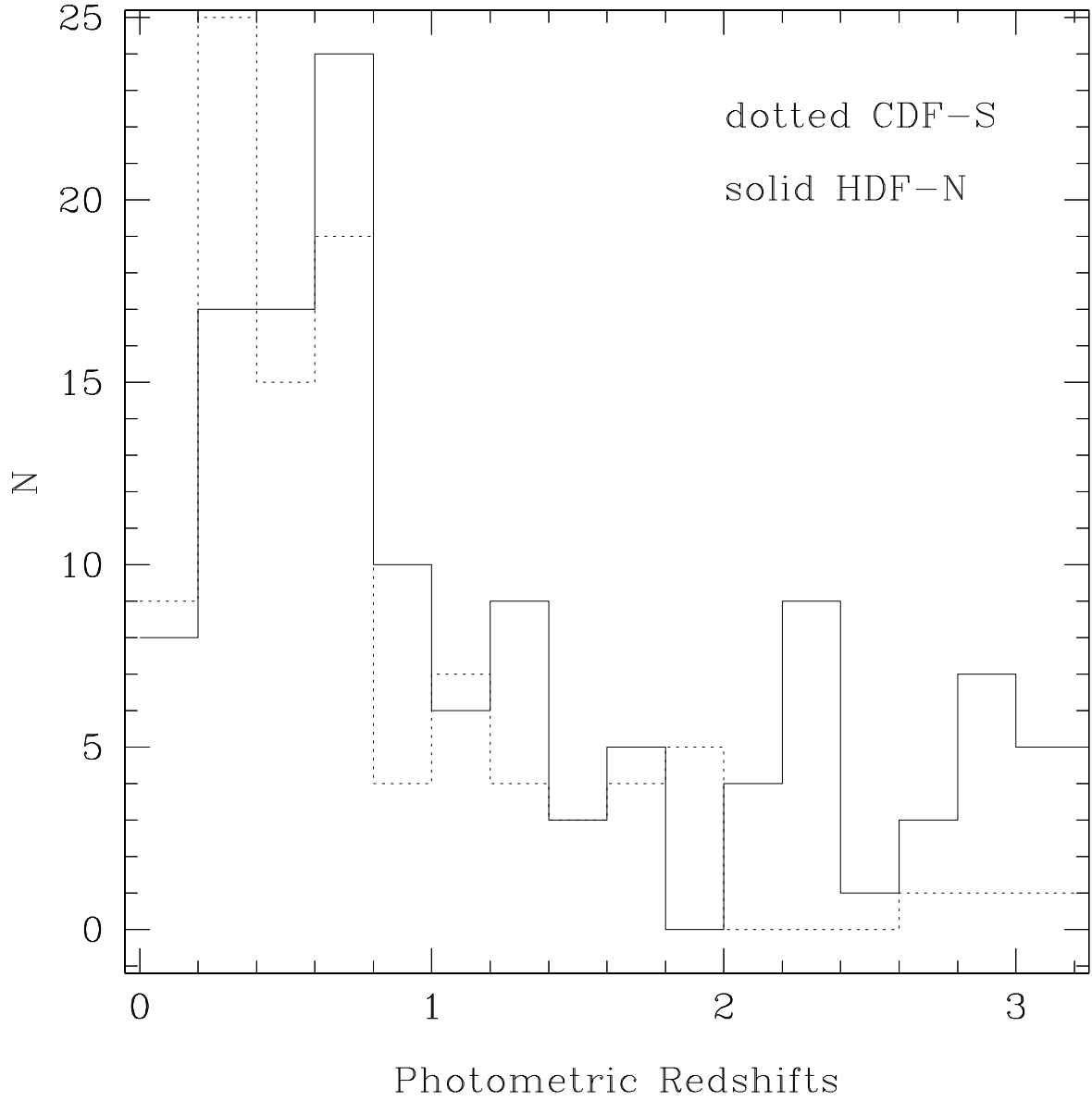}
\caption{Photometric redshift distribution for the northern (solid line) and southern (dotted line) 
fields. \label{histpz}}
\end{figure}

\begin{figure}
\plotone{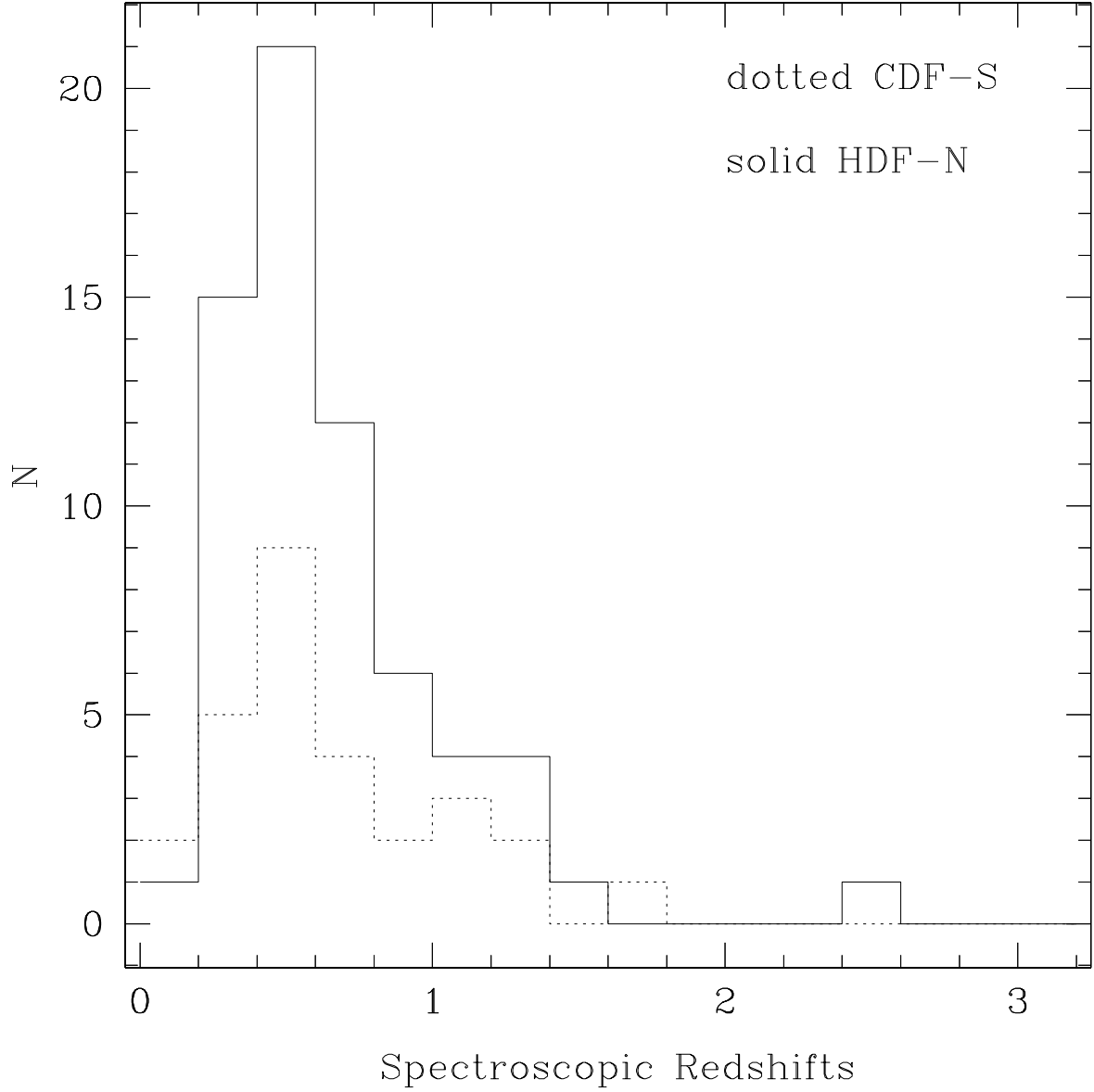}
\caption{Spectroscopic redshift distribution for the northern (solid line) and southern (dotted line) 
fields. \label{histz}}
\end{figure}

\begin{figure}
\plotone{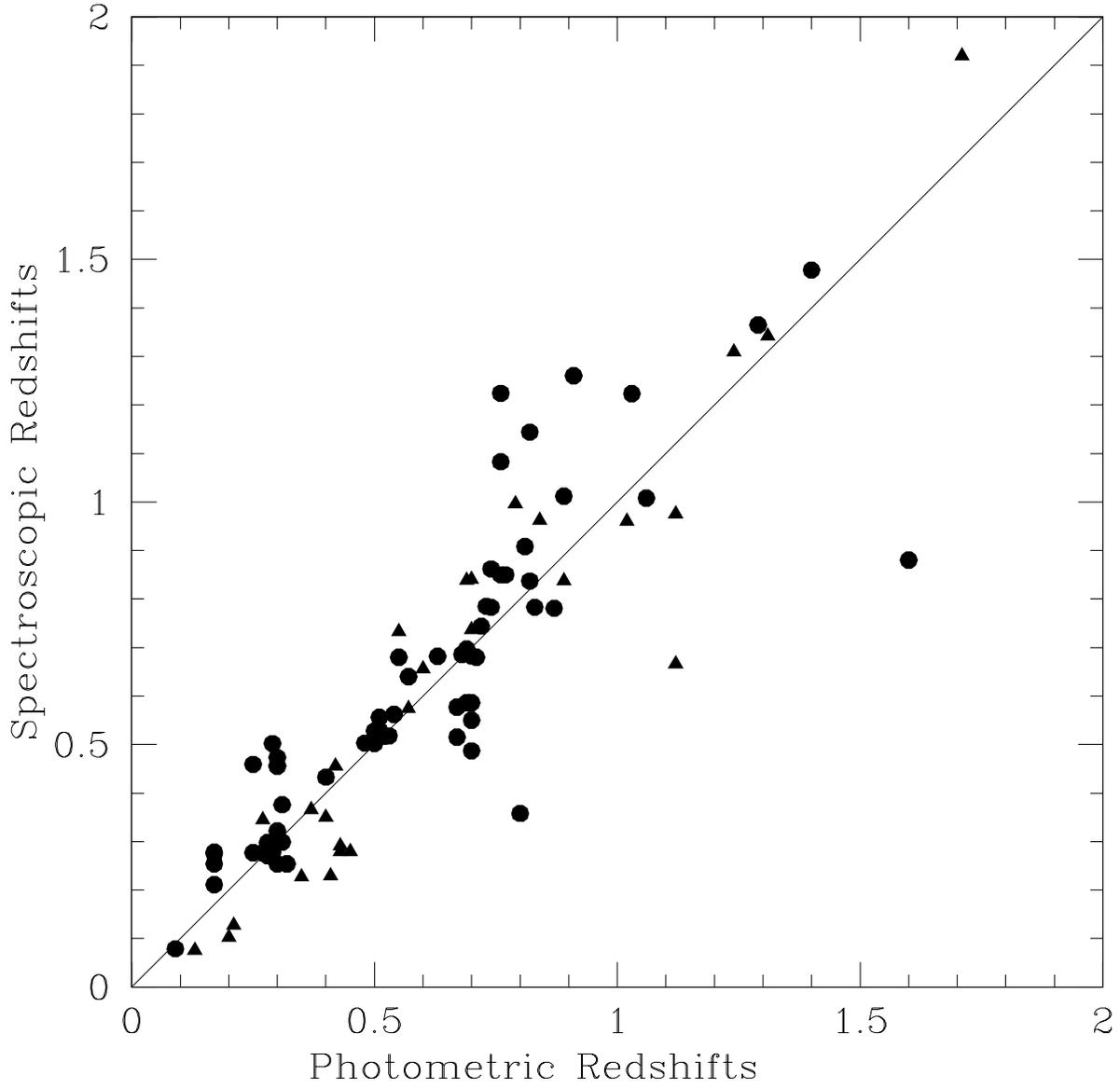}
\caption{Photometric and spectroscopic redshifts for the northern (circles) and southern (triangles) 
fields. \label{phtzspecz}}
\end{figure}

Spectroscopic redshifts of 95 objects of the UV-selected sample, 30 in the southern field and 65 in the northern
field, are available through the GOODS collaborations 
(taken from the ESO/GOODS-CDFS spectroscopy master
catalog\footnote{
http://www.eso.org/science/goods/spectroscopy/CDFS$_{-}$Mastercat/} and from the Team Keck list by Wirth et al. 2005). 
Table 1 shows the magnitudes U, B, V, i, z and errors, redshifts (photometric and spectroscopic when available),
and spectral types for a few galaxies of our sample. The entire sample 
will be made available in the eletronic version of this paper. 
Fig.~\ref{histpz} and Fig.~\ref{histz} show the photometric and spectroscopic redshift 
distributions for both fields. The agreement between spectroscopic
and photometric redshifts as shown in Fig.~\ref{phtzspecz} is within the redshift accuracy
of GOODS (Dahlen et al. 2005). We have checked the four objects with the most discrepant redshifts and concluded that
the main reason for the discrepancy is due to either the presence of bright companions which 
makes the photometry of the objects less accurate, or to confusion on the slit when the
spectra were taken. 

\begin{figure}
\plotone{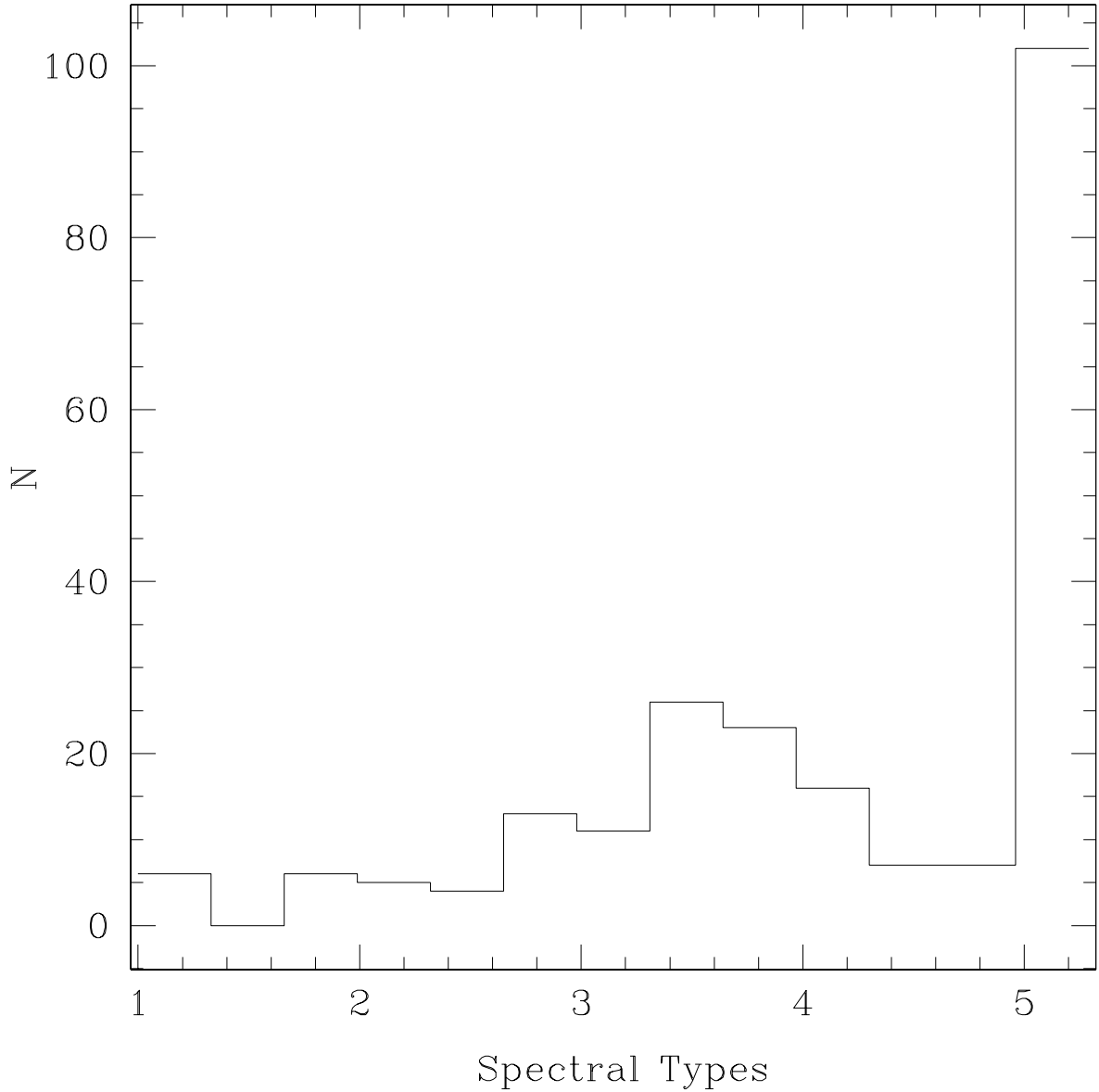}
\caption{Spectral Type distribution for the UV-selected galaxies in the two GOODS fields. 
E (1), Sbc (2), Scd (3) and Im (4) are from Coleman et al. (1980). Spectral type (5) corresponds
to either of the two starburst templates from Kinney et al. (1996).
\label{histst}}
\end{figure}

The distribution of the spectral types (ST) which were obtained from the template fitting when deriving the  
photometric redshifts are shown in Fig.~\ref{histst}. As expected in the case of UV-selected samples,  
the majority ($\sim$48\%) of the objects are starbursts (ST$>$4.5). In fact, the fraction of starbursts 
could be even higher since a starburst galaxy with a high internal extinction has a redder SED and may be
shifted to a earlier spectral type with ST$<$4.5. A few galaxies with 
early-type SEDs ($\sim$5\% have ST$<$1.5) are also present in the UV-selected sample. These early-type objects 
with star-forming cores could be the result of mergers or star formation triggered by the rapid
mass infall into a central black hole as suggested in Menanteau et al. (2001). 
For comparison, Burgarella et al. (2006) found that LBGs at $z \sim 1$ are
either disks (75\%) or mergers (22\%), and only one object (3\%) is a spheroid.

\section{Starbursts} 

In order to assess the nature of the galaxies that are forming stars at intermediate$-z$ 
we have visually inspected all 93 objects with starbursts SEDs and
classified them according to their optical morphology and searched their environment (5$\times$5 arcsec)
for other objects. We did the analysis in the GOODS bands (B, V, i, and z) since FUV morphologies can give an erroneous view of the galaxy's shape (e.g. Hibbard \& Vacca 1997). 
We found that (i) nearly 75\% of the starbursts have tidal tails or show some peculiarity 
typical of interaction/mergers, (ii) $\sim$50\% have another galaxy 
within the searched area. 

We also found that $\sim$36\% of the starbursts are characterized 
by the presence of a large clump and extended tail, known as tadpoles. Very little is known about
these objects and their fate during evolution. The main question is whether such galaxies are 
rotating edge-on systems or are objects with peculiar morphology from merging or accretion 
(Elmegreen et al. 2004, 2005a). We discuss these objects in more detail in \S 6.

Although our photometric
redshifts are well-calibrated with spectroscopic redshifts within the GOODS
collaboration ($>$1,000 redshifts for both GOODS fields), the featureless spectra of starbursts,
compared to earlier type galaxies with a pronounced 4000\AA-break, make their
photometric redshift estimates uncertain with an increased
risk of `catastrophic redshifts' with large errors. In order to prevent any bias due to
the photometric redshift estimates we were conservative and repeated our analysis for 
the 95 objects with known spectroscopic redshifts and found (i) 35\% are starbursts, 
(ii) 66\% of the starbursts show some obvious tidal effect and/or presence of another galaxy 
within the searched area. Only 33\% of the other spectral types show tidal tails and/or presence 
of other galaxies within the same area. 

\begin{figure*}
\plotone{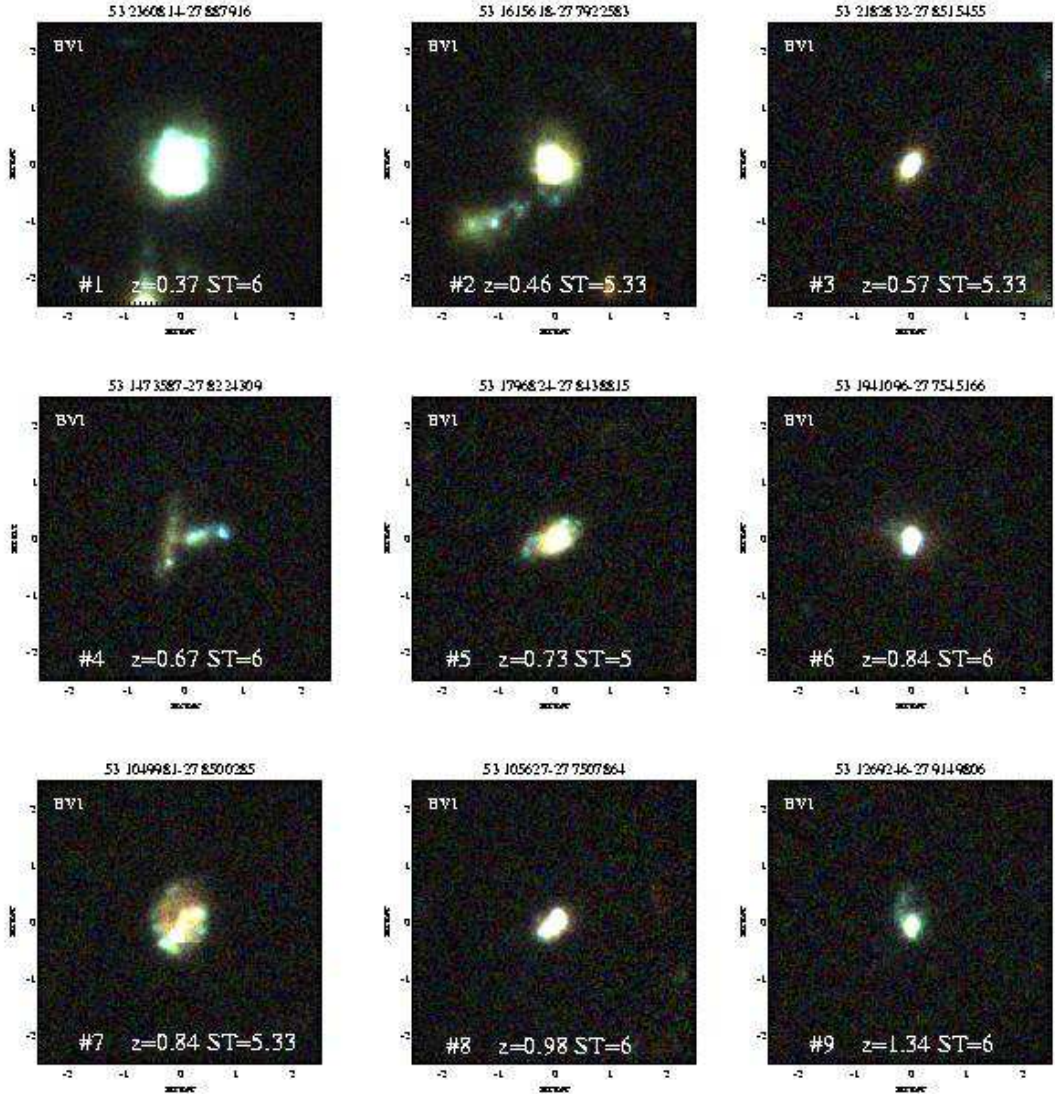}
\caption{Starbursts in the CDF-S, spectroscopic redshifts and spectral types are shown
together with a 3-color (BVi) postage stamp montage of 
5$\times$5 arcsec$^{2}$.\label{starbursts}}
\end{figure*}

Fig.~\ref{starbursts} shows a gallery of starbursts from the CDF-S with spectroscopic
redshifts at $z=0.37-1.34$. Galaxies of all types are represented in the starburst gallery.
For example, an early-type (\#3) and a merging system (\#4), disk galaxies with strong knots of 
star-formation (\#1, \#5 and \#7), compact objects or clumps with tidal tails (\#6 and \#9). 
Strong interacting systems such as \#1 and \#2 are also common in the UV-selected sample of 
starbursts.

\section{Discussion}

\begin{figure}
\plotone{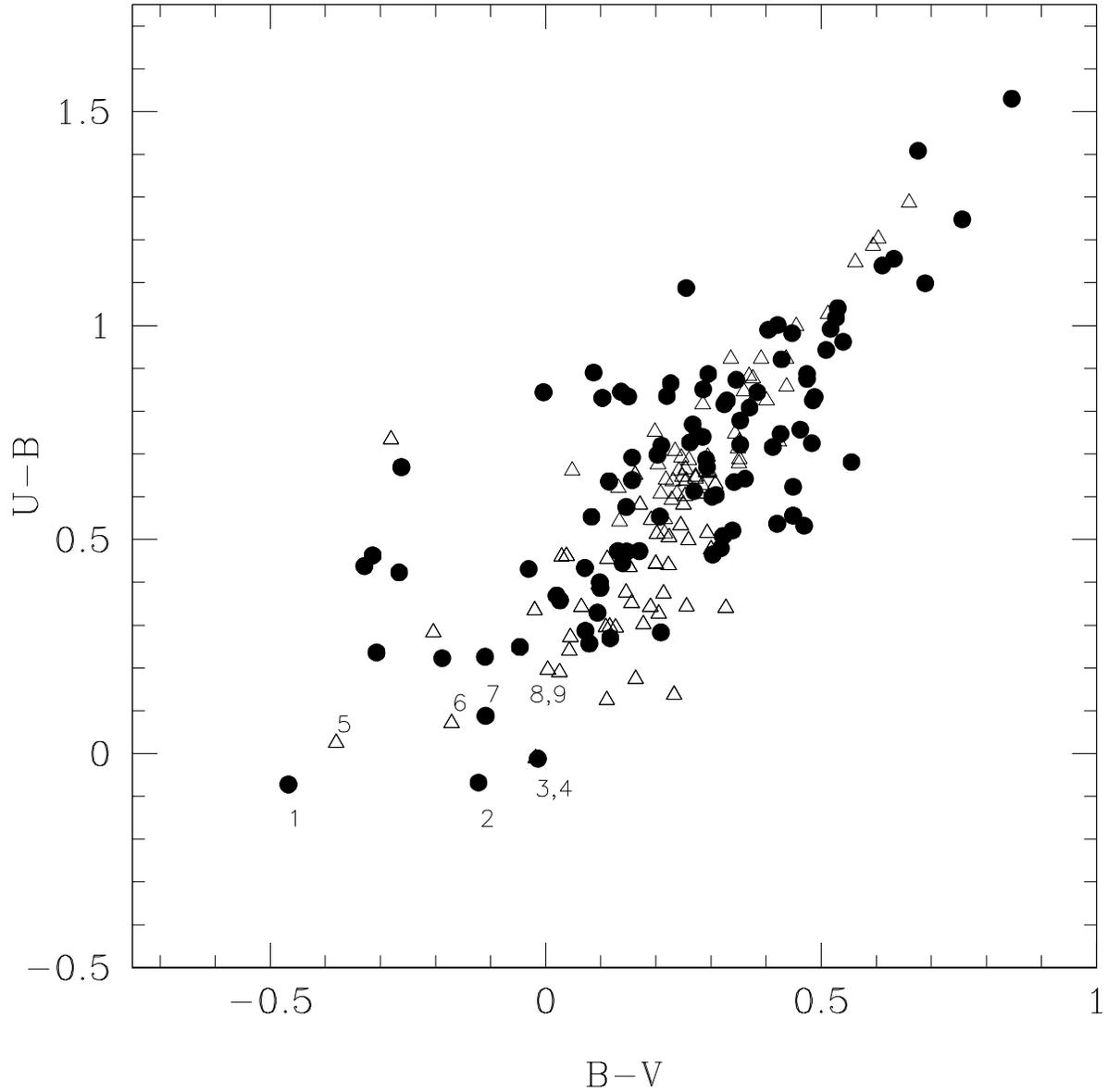}
\caption{Rest-frame U-B versus B-V color. Galaxies with spectral types $>$ 4.5 (i.e. starbursts) are marked
in blue. The northern field objects are in circles and the southern field in triangles.  
Only galaxies with $z_{\rm phot}$ $<2$ were included. Bluest objects are identified by numbers 1--9.
\label{plotubbvpz2}}
\end{figure}

Using information from the photometric redshifts or spectroscopic when available, rest-frame absolute 
magnitudes and colors are calculated using the recipe in Dahlen et al. (2005). 
However, we only analyzed the ones with $z_{\rm phot}$$< 2$ since 
photometric redshift uncertainties are larger for higher redshifts. 
Fig.~\ref{plotubbvpz2} shows the rest-frame U--B versus B--V color for both fields. The  bluest objects 
($U-B  < 0.2$ and $B-V < 0.1$) have spectral types of starbursts, except for one object (\#2 in Fig.~\ref{plotubbvpz2})
with spectral type $Im$, which is also star-forming and UV-bright.

\begin{figure}
\plotone{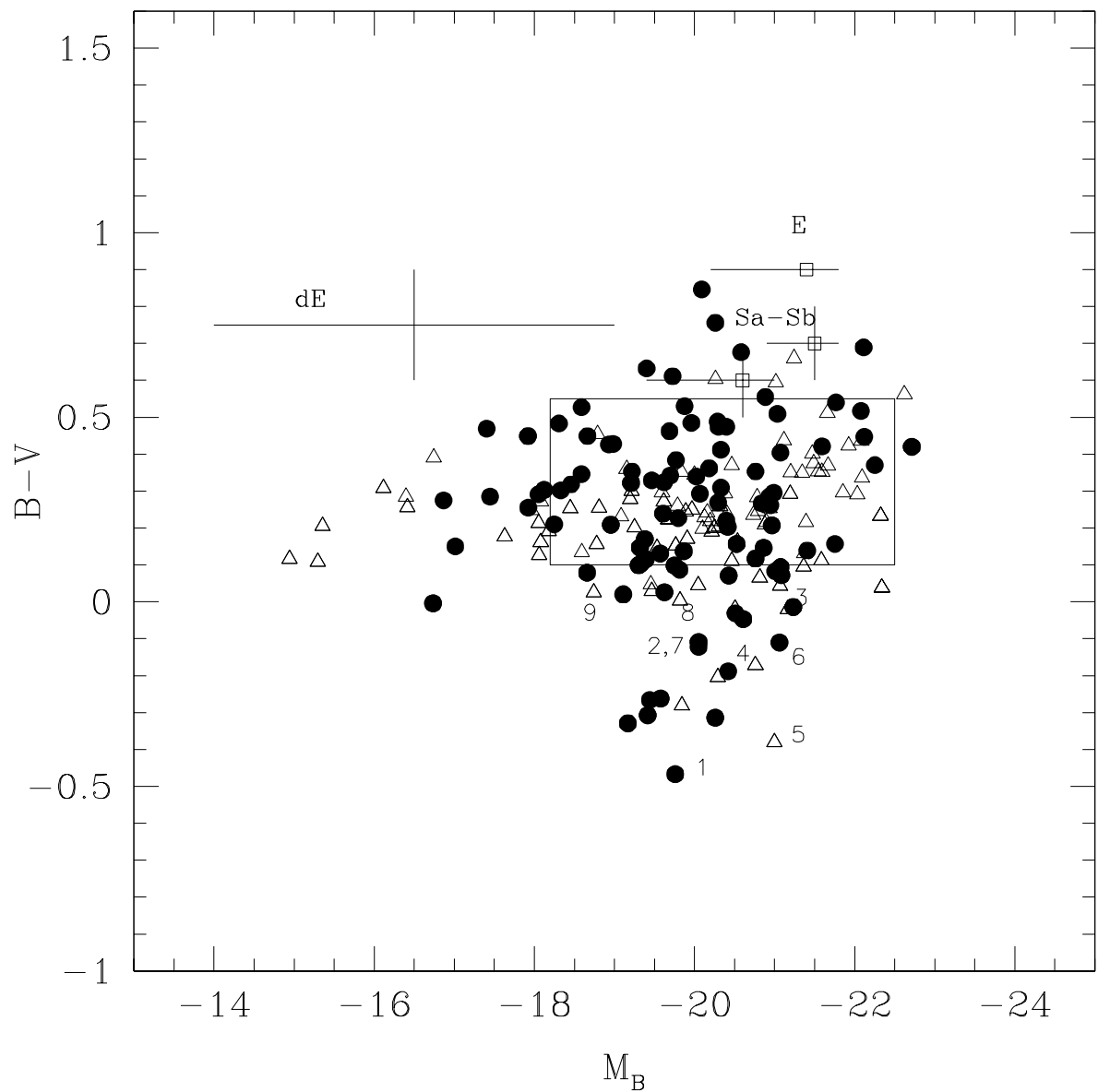}
\caption{Rest-frame magnitude B versus B-V color. Galaxies with spectral types $>$ 4.5 (i.e. starbursts) are marked
in blue. The northern field objects are in circles and the southern field in triangles. 
The three squares are values typical of E, Sa-Sb, Sc-Irr (clockwise); the cross on the top 
left corresponds to the dE and dSph; the box region corresponds to the strong star-forming 
galaxies (Bershady et al. 2000) which contains blue nucleated galaxies, compact narrow emission-line 
galaxies and small, blue galaxies at intermediate redshifts. Only galaxies with $z_{\rm phot}$ $<2$ were
included. Numbers 1--9 are the same objects as in Fig.~\ref{plotubbvpz2}.
\label{plotbbvpz2}}
\end{figure}

The bluest objects are also marked in the color-magnitude plot shown in 
Fig.~\ref{plotbbvpz2} where we have also identified 
objects from Bershady et al. (2000) which includes typical Hubble types, dwarf 
ellipticals and luminous blue compact galaxies at intermediate redshifts.
A large number of the UV-selected sample fits the latter category which Bershady et al. 
described as blue nucleated galaxies, compact narrow emission-line galaxies and small, blue 
galaxies at intermediate redshifts. A similar color-magnitude plot was presented in de Mello et al. (2005, their Fig.~13)
in the analysis of the deepest U--band image ever taken with HST. The main difference between their 
color-magnitude distribution and
Fig.~\ref{plotbbvpz2} is that the shallow U--band survey, presented here, detected a larger number of brighter and bluer
objects than the deep survey. This is expected, since the wide survey covers a much larger area (the
deep survey is only slightly larger than one WFPC2 field) and is sensitive only to bright galaxies.  
The average rest-frame M$_{\rm B}$ value for the shallow U--band  (M$_{\rm B}$$= -19.91 \pm 0.10$) is one magnitude 
brighter than the average value of the deep U--band (M$_{\rm B}$$= -18.43 \pm 0.13$). 
Despite the fact that the shallow U--band covers a larger area and is sensitive only to the bright 
end of the luminosity distribution, none of the UV-selected galaxies in the GOODS fields is brighter than M$_{\rm B}$= --23.

\begin{figure*}
\plotone{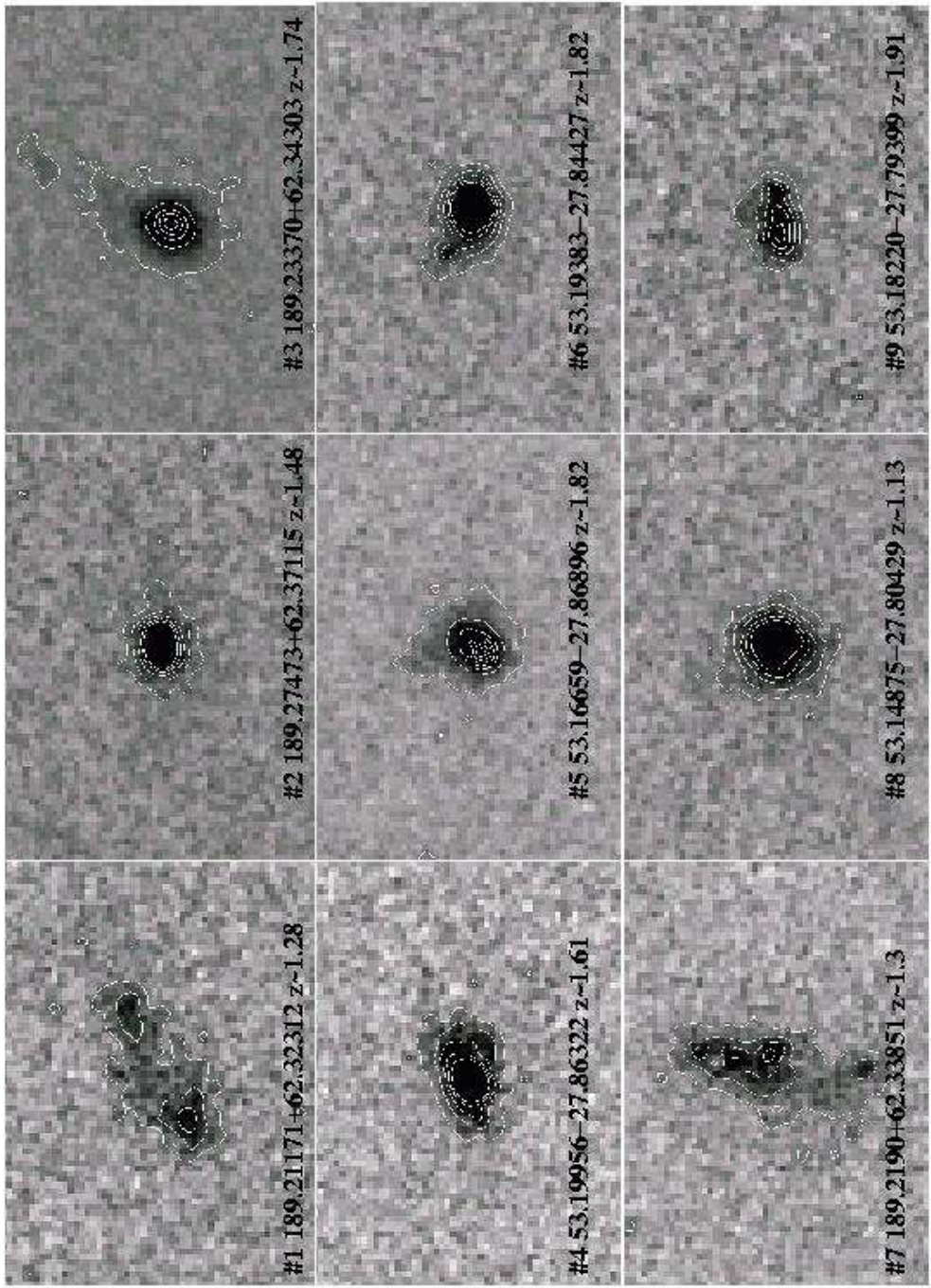}
\caption{Contours on top of z--band images of the bluest objects in the 
U--B versus B--V plot shown in Fig.~\ref{plotubbvpz2}. 
All objects have spectral types of starbursts, except for object \#2 which has spectral type of Im galaxies. Contours
limits are mininum=0.002, maximum=0.01, 5 levels, except for object \#3 and \#5 which have maximum
values 0.1 and 0.05 and 9 levels, respectively. Size of each image is 76 $\times$ 53 pixels.
\label{bluetail}}
\end{figure*}

The bluest objects are at $z=1.1-1.9$ and as shown by the contours in Fig.~\ref{bluetail} have peculiar 
morphologies that resembles either tadpoles, chains, or double-clump galaxies. 
These galaxies are characterized by the presence of a large clump and extended tail, 
or two or more large clumps in a linear arrangement. Like the objects in the UDF and 
the Tadpole ACS field analyzed by Elmegreen et al. (2004, 2005a), their unusual
morphology is not merely a band-shifting effect, since the large clumps 
have no counterparts in the local Universe as seen in the UV nearby galaxy survey 
(Windhorst et al. 2002). The main question
is whether these clumps are accreted clumps that are building up galaxies through 
hierarchical mergers (Straughn et al. 2004, Elmegreen et al. 2004 and Elmegreen et al. 2005b)
or the result of gravitational instabilities of gas accreting in a turbulent
medium in a disk (e.g., Noguchi 1996; Immeli 2004). 
These questions could be answered in the future with near-IR spectroscopy.

\begin{figure*}
\plotone{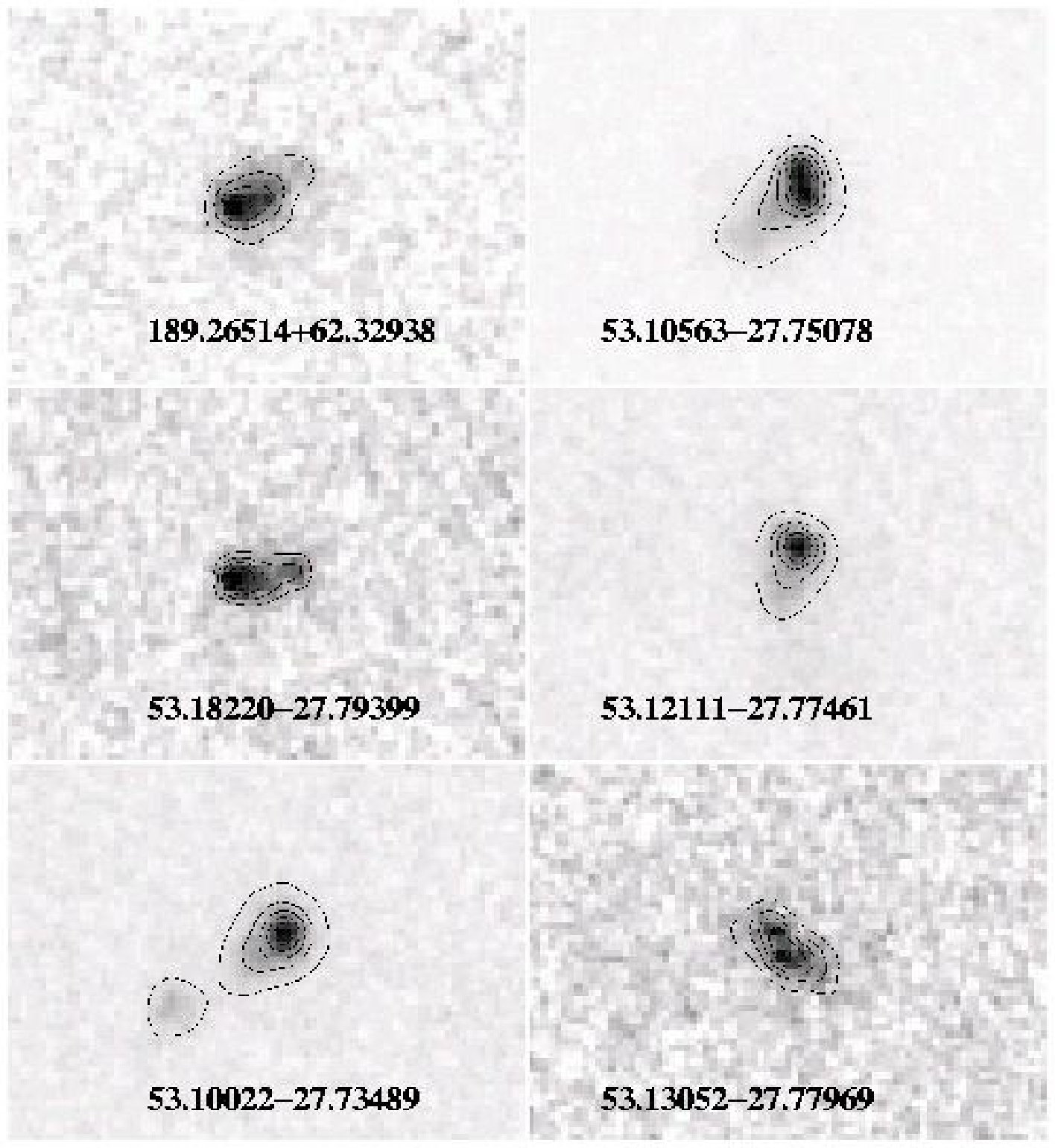}
\caption{Contours on top of B--band image of the tadpole starbursts at $z\sim 1$. Coordinates are given for
each object.
\label{contours1}}
\end{figure*}

\begin{figure*}
\plotone{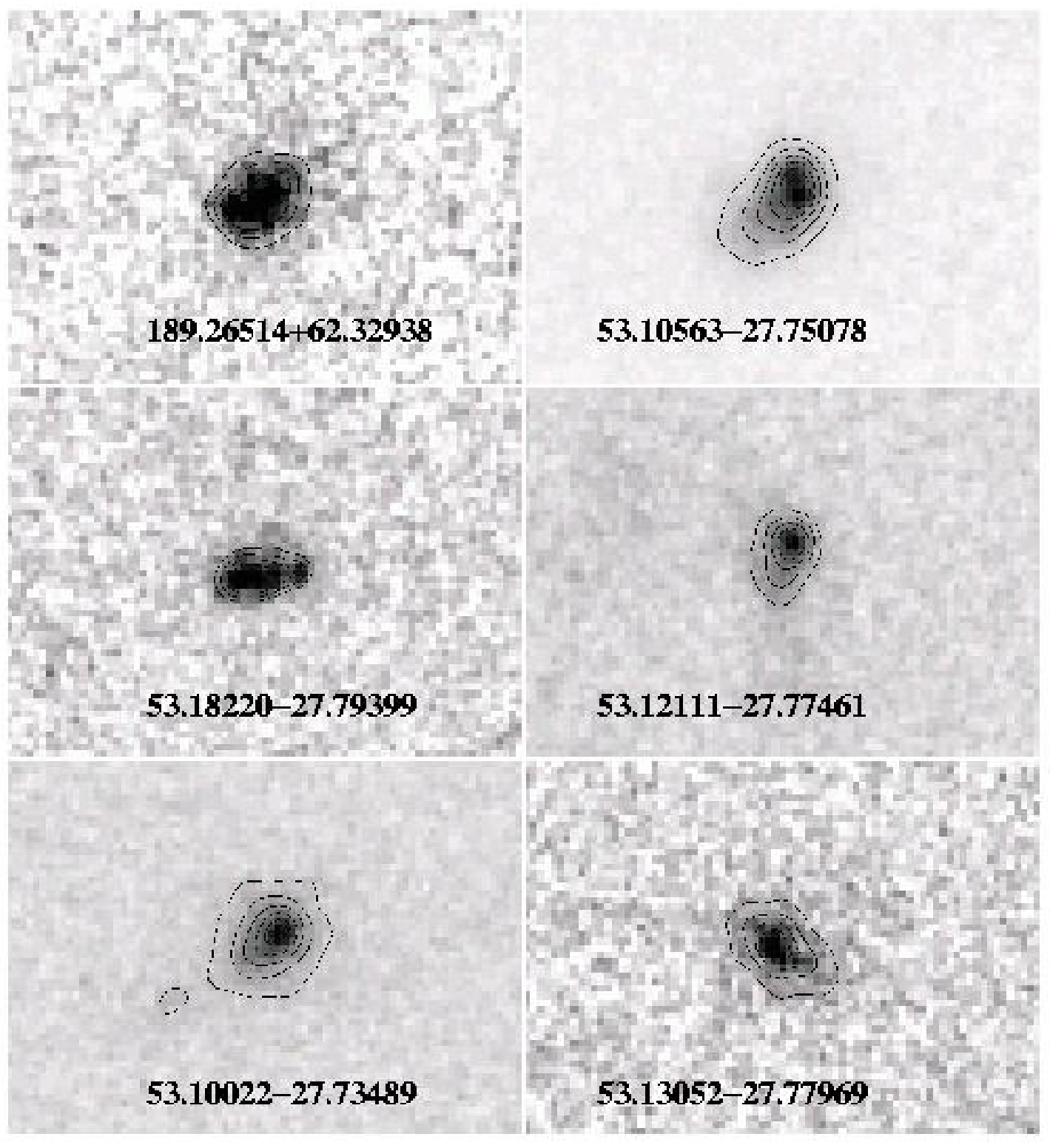}
\caption{Contours on top of z--band image of the tadpole starbursts at $z\sim 1$. Coordinates are given for each object.
\label{contours1z}}
\end{figure*}

There are twelve galaxies in our sample in the redshift
range $z=0.8-1.3$ and they are either interacting systems, such as object \#4 in the starburst
gallery, or have the tadpole/clump morphology discussed above. 
Fig.~\ref{contours1} and \ref{contours1z} show six of these 
objects in the B and z band. At $z \sim 1$ the B--band is showing the rest-frame near-UV morphology while the z--band is showing
the morphology in the rest-frame B-band. The contours show that the morphologies are similar
in both bands.  
The average total magnitudes and half-light radii of these objects are M$_{\rm B}$=$-20.21\pm1.11$ 
and $1.63 \pm 0.37$ kpc (measured in the B--band), respectively. These starbursts 
at $z=0.8-1.3$ have sizes comparable to LBGs and compact UVLGs. 

\section{Summary}

We present a sample of 268 objects, including 42 stars or AGNs,  in the GOODS North and South fields which
was selected from `shallow' images taken with the U--band filter (F300W) during the 
WFPC2 parallel observations of the GOODS ACS campaign. The analysis of the
ACS images (B, V, i, z) of these UV-selected objects reveals that: 

\begin{enumerate}

\item Most of the objects are between $0.2<z<0.8$,
\item The majority (45\%) of the galaxies have spectral types of starbursts, however,
galaxies of all spectral types are found, including early-types (5\%), 
\item 75\% of the starbursts have tidal tails or show some peculiarity typical of 
interaction/mergers and  50\% of the starbursts have another galaxy within 5$\times$5 arcsec; 
\item The bluest galaxies ($U-B<0.2$ and $B-V<0.1$) are at $1.1<z<1.9$ and have peculiar 
morphologies that resembles either tadpoles, chains, or double-clump galaxies. 
\item Starbursts at $z \sim 1$ with tadpole/clumps morphology have half-light radius of $1.6 \pm 0.4$kpc.

\end{enumerate}

\acknowledgments

We are grateful to the GOODS team. Support for this work was provided by NASA through grants 
GO09583.01-96A and GO09481.01-A from the Space Telescope Science Institute, which is operated 
by the Association of Universities for Research in Astronomy, Inc., under NASA contract NAS5-26555.

\clearpage

\begin{deluxetable}{ccrccccccccccccc}
\tabletypesize{\tiny}
\tablecaption{CDF-S Near-UV sources - UBViz}
\tablewidth{0pt}
\tablehead{
\colhead{RA } & 
\colhead{Dec  } & 
\colhead{Exp} &
\colhead{u	      } & 
\colhead{u$_{\rm err}$    } & 
\colhead{b } & 
\colhead{b$_{\rm err}$	       } & 
\colhead{v	    } & 
\colhead{v$_{\rm err}$		    } & 
\colhead{i	    } & 
\colhead{i$_{\rm err}$		    } & 
\colhead{z	    } & 
\colhead{z$_{\rm err}$		    } &
\colhead{$z_{\rm phot}$		    } &
\colhead{ST	       } &
\colhead{$z_{\rm spec}$} 
}
\startdata
 53.1025057 & -27.7411446 & 800 & 23.38   & 0.05 & 23.30 & 0.04 & 22.61 & 0.02 & 22.03 & 0.02 & 21.86 & 0.02 & 0.51 & 3.7 & 0.54 \\  
 53.1056270 & -27.7507864 & 2400& 23.48   & 0.05 & 23.09 & 0.03 & 23.12 & 0.02 & 22.69 & 0.02 & 22.51 & 0.02 & 1.12 & 6.0   & 0.97 \\  
 53.1258995 & -27.7512749 & 2260& 23.17   & 0.02 & 22.55 & 0.02 & 22.03 & 0.01 & 21.62 & 0.01 & 21.40 & 0.01 & 0.43 & 5.3 & 0.29 \\  
 53.1237751 & -27.7520043 & 2260& 23.92   & 0.06 & 23.60 & 0.03 & 23.10 & 0.02 & 22.39 & 0.02 & 22.20 & 0.02 & 0.70 & 3.7 & 0.74 \\  
 53.1952443 & -27.7537776 & 1500& 22.72   & 0.02 & 23.96 & 0.04 & 23.26 & 0.02 & 22.26 & 0.02 & 21.82 & 0.02 & 0.89 & 2.7 & 0.84 \\ 
\enddata				                      	     												    
\end{deluxetable}

\end{document}